\documentclass[longbibliography,twocolumn,prl,aps,superscriptaddress,showpacs,amsmath,amssymb,floatfix]{revtex4-1}
\usepackage{graphicx}
\usepackage{amssymb}
\usepackage{amsmath}
\usepackage{epsfig}
\usepackage{color}
\usepackage{mathtools}
\usepackage[colorlinks,linkcolor=blue,anchorcolor=blue,citecolor=blue,urlcolor=blue]{hyperref}
\usepackage{physics}
\usepackage{bm}
\usepackage{tikz}
\usepackage{feynmf}
\usepackage[compat=1.1.0]{tikz-feynman}
\usepackage{multirow}
\usepackage[normalem]{ulem}

\begin{document}
	
	\title{Quantumness and quantum to classical transition in the generalized Rabi model} 
	\author{Wei-Feng Zhuang}
	\email{zhuangwf@baqis.ac.cn}
	\affiliation{CAS Key Laboratory of Quantum Information, University of Science and Technology of China, Hefei, 230026, China}
	\affiliation{Beijing Academy of Quantum Information Sciences, Beijing 100193,China}
	\author{Yun-Tong Yang}
	\affiliation{School of Physical Science and Technology, Lanzhou University, Lanzhou 730000, China}
	\affiliation{Lanzhou Center for Theoretical Physics \& Key Laboratory of Theoretical Physics of Gansu Province, Lanzhou University, Lanzhou 730000, China}
	\author{Hong-Gang Luo}
	\email{luohg@lzu.edu.cn}
	\affiliation{School of Physical Science and Technology, Lanzhou University, Lanzhou 730000, China}
    \affiliation{Lanzhou Center for Theoretical Physics \& Key Laboratory of Theoretical Physics of Gansu Province, Lanzhou University, Lanzhou 730000, China}
	\affiliation{Beijing Computational Science Research Center, Beijing 100084, China}
	\author{Ming Gong}
	\email{gongm@ustc.edu.cn}
	\affiliation{CAS Key Laboratory of Quantum Information, University of Science and Technology of China, Hefei, 230026, China}
\affiliation{Synergetic Innovation Center of Quantum Information and Quantum Physics, University of Science and Technology of China, Hefei, Anhui 230026, China}
\affiliation{CAS Center For Excellence in Quantum Information and Quantum Physics,  University of Science and Technology of China, Hefei, Anhui 230026, China}
	\author{Guang-Can Guo}
	\affiliation{CAS Key Laboratory of Quantum Information, University of Science and Technology of China, Hefei, 230026, China}
\affiliation{Synergetic Innovation Center of Quantum Information and Quantum Physics, University of Science and Technology of China, Hefei, Anhui 230026, China}
\affiliation{CAS Center For Excellence in Quantum Information and Quantum Physics,  University of Science and Technology of China, Hefei, Anhui 230026, China}
	\date{\today }
	
	\begin{abstract}
	The quantum to classical transition (QCT) is one of the central mysteries in quantum physics. This process is generally interpreted as state collapse from measurement or decoherence from interacting with the environment. 
	Here we define the quantumness of a Hamiltonian by the free energy difference between its quantum and classical descriptions, which vanishes during QCT. We apply this criterion to the many-body Rabi model and study its scaling law across the phase transition, finding that not only the temperature and Planck constant, but also all the model parameters are important for this transition. We show that the Jaynes-Cummings and anti Jaynes-Cummings models exhibit greater quantumness than the Rabi model. Moreover, we show that the rotating wave and anti-rotating wave terms in this model have opposite quantumness in QCT. 
	We demonstrate that the quantumness may be enhanced or suppressed at the critical point. 
	Finally, we estimate the quantumness of the Rabi model in current trapped ion experiments. 
	The quantumness provides an important tool to characterize the QCT in a vast number of many-body models.
	\end{abstract}
	
	\maketitle
	Quantum and classical physics differ fundamentally. In quantum mechanics, particles are 
    described as waves, exhibiting phenomena
	 such as entanglement and superposition. 
	In contrast, classical physics precludes these characteristics, with well-defined mass points satisfying 
	the Newton's laws. The quantum to classical transition (QCT) is explained through the orthodox Copenhagen 
	interpretation of wave function collapse from measurement 
	\cite{stapp1972copenhagen,
	bassi2013models,cramer1986transactional,camilleri2015niels},
	which is then understood by the quantum decoherence theory 
	\cite{zurek1991decoherence, Haroche_1998, Zeh1970, paz2002environment,schlosshauer2014quantum,
	zeh1970interpretation,zurek1981pointer,schlosshauer2005decoherence,zurek1982environment,zurek2003decoherence}. 
	The QCT is a rather subtle issue. While in principle no such boundary exists, because the bulk 
	metals and superconductors in solids and the neutron star in astrophysics should be described by 
	quantum mechanics; however, there are also 
	abundant evidences that when the system temperature is high enough, or the Planck constant 
	becomes negligible, the quantum nature may disappear. Furthermore, the importance of system size can be
	evidenced from macroscopic Schr\"odinger cat \cite{Haroche_1998,leong2020large,lewenstein2021generation,wang2022flying,bild2023schrodinger}. Therefore, 
	these factors should be among the most crucial factors for this QCT. Unfortunately, a quantitative 
	characterization of this transition is still 
lacking. 
	
	{\it Definition of quantumness}: Imagine a Hamiltonian $\mathcal{H}(x, p)$, where $x$ and $p$ are coordinate and 
	momentum operators, respectively. In classical physics, $xp = px$ are the same; yet in quantum physics, 
	they are different, with $[x, p] = i\hbar$, where $\hbar$ is the Planck constant divided by $2\pi$. 
	Obviously, in quantum mechanics, $x^2$ and $p^{-1} x p^2 x p^{-1}$ are different. 
	Thus, when we say that a system exhibits quantum behaviors, we mean that it possesses some features 
	quantitatively different from the classical counterpart. Since in thermodynamics all observations can 
	be determined by the partition function, we expect their difference can be reflected from their free energies. 
	When these two descriptions yield the same free energy, they can not be distinguished from measurements. 
	Based on this intuitive picture, we define the quantumness as 
	\begin{equation}
		\Delta_\text{QC} = F_\text{Q} - F_\text{C},
		\label{eq-def}
	\end{equation}
	where $F_\text{Q}$ and $F_\text{C}$ are the free energies using quantum and classical mechanics. 
	The QCT is exact when $\Delta_\text{QC} = 0$, and the nature of quantum vanishes. 
	This work examines this quality in the many-body Rabi model, which hosts quantum phase transition. 
	Strikingly, we show that the critical point with significant quantum (or classical) fluctuation is not 
	necessarily associated with maximal quantumness. Finally, we will estimate the role of quantumness in 
	the current trapped ion experiments.
	
	We can gain some insight into this problem by investigating some simple models.
	(A) A free particle model with $\mathcal{H} = p^2/2m$, where $m$ is the particle mass. The partition functions
	are $Z_\text{Q} = Z_\text{C} = \int dx dp/(2\pi \hbar) e^{-\beta \mathcal{H}} 
	=V \sqrt{2m\pi /\beta} $, where $V$ is the real-space volume, and $\beta = 1/k_BT$, with $k_B$ is the 
	Boltzmann constant and $T$ is the temperature. Hence $\Delta_\text{QC} = 0$.  
	(B) A quantum oscillator model with $ \mathcal{H} = p^2/2m + m \omega^2 x^2/2$, with eigenvalues 
	$E_n = \hbar\omega(n+1/2)$. We have $Z_\text{Q} = \sum_{n=0}^\infty e^{-\beta \hbar \omega(n+1/2)}$, and $Z_\text{C} = 
	\int dx dp/(2\pi\hbar) e^{-\beta \mathcal{H}}$, yielding 
	\begin{equation}
		\Delta_\text{QC} = {1\over 24} (\hbar\omega)^2 \beta - \frac{1}{2880}(\hbar \omega)^4 \beta^3  + \cdots,
		\label{eq-harmonic}
	\end{equation}
	when $\beta \hbar \omega \ll 1$, which is always a positive value. 
	When $\omega = 0$, it reduces to model (A). Thus when $k_B T \gg \hbar\omega$,
	the quantization effect is not important, and QCT happens. In the presence of 
	many-body interaction, the level spacing is increased and the quantumness $\Delta_\text{QC}$ will 
	also be increased in accordingly. (C) QCT from their statistics. One can find that the crossover between Bose-Einstein 
	and Fermi-Dirac distributions and the Boltzmann-Maxwell
	distribution can be reached when $\beta \mu \ll -1$ ($\mu$ is the chemical potential). The last two models 
	show that the system parameters are also important for QCT. 
	
	\begin{figure}
		\includegraphics[width=0.4\textwidth]{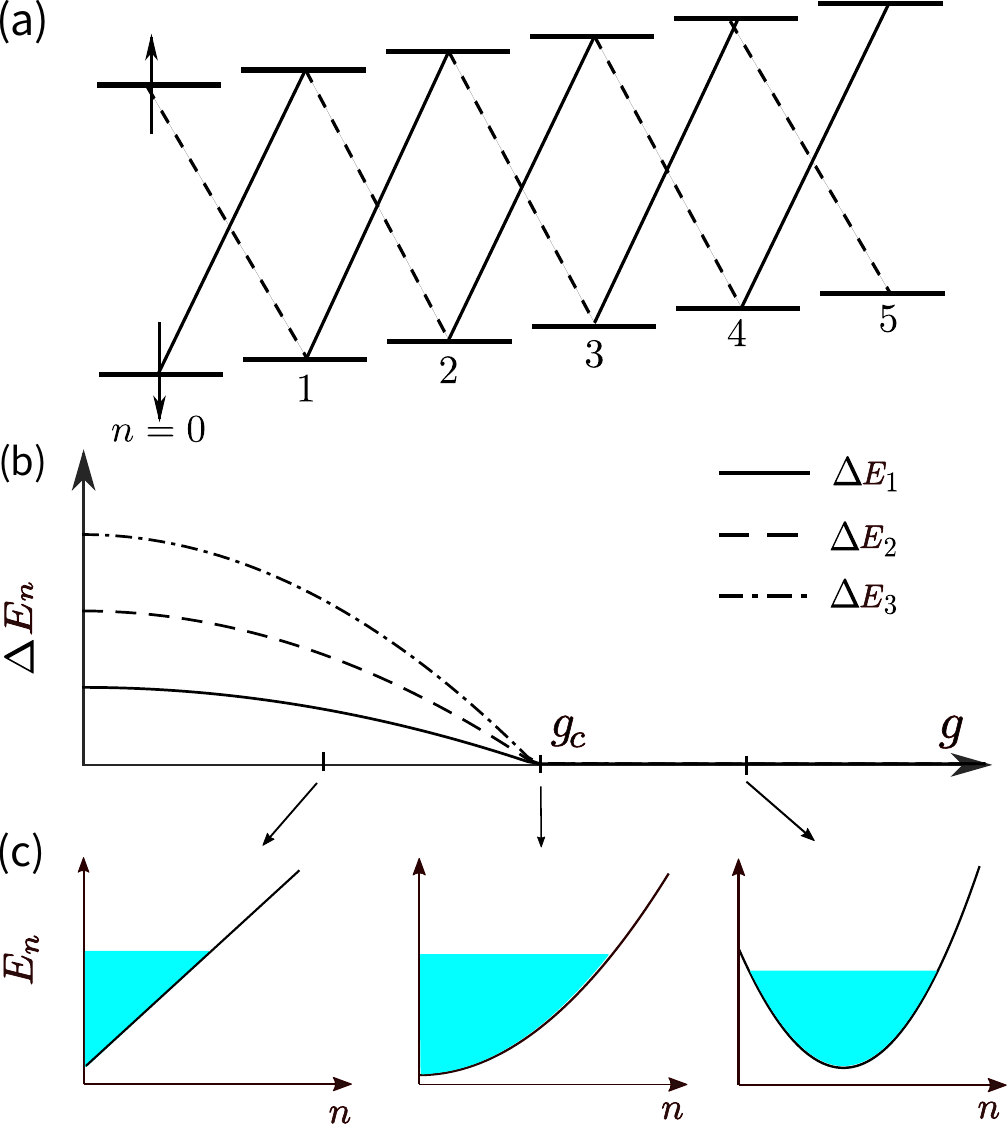}
		\caption{(a) Energy levels of the generalized Rabi model, in which the solid and dashed lines correspond to coupling induced by
			rotating wave term ($g_1$) and anti-rotating wave term ($g_2$), respectively. When $g_2 = 0$ (JC model) 
			or $g_1 = 0$ (aJC model), this model can be solved exactly; see discussions in (I) 
			and (II). (b) Phase transition in this model characterized by energy gaps $\delta E_n = E_n  -E_0$ by 
			tuning of $g_i$. (c) Dispersion of energy levels in the normal
			phase ($g < g_\text{c}$), critical point (at $g_\text{c}$) and superradiant phase ($g > g_\text{c}$) when $\omega/\Omega
			\rightarrow 0$.}
		\label{fig-fig1}
	\end{figure}
	
	{\it Quantumness in the generalized Rabi model}:
	With these results in mind, we examine the physics in the generalized Rabi model 
	\cite{li_quantum_2006,alcalde_path_2011,chen_exact_2012, moroz2013solvability, batchelor2015integrability, bakemeier_quantum_2012, Braak_2016, wolf_dynamical_2013,braak2011integrability}
	\begin{equation}
		\mathcal{H} = \hbar \omega a^\dagger a + {\hbar \Omega \over 2} \sigma^z + \hbar g_1 a \sigma^\dagger + 
		\hbar g_2 a \sigma^- + \text{h.c.},
		\label{eq-H}
	\end{equation}
	where $\omega$ is the harmonic oscillator frequency, $\Omega$ is the spin level spacing, 
	and $g_1$, $g_2$ are the couplings of rotating and anti-rotating wave terms, respectively. 
	This model has been intensively studied using trapped ions \cite{lv2018quantum,puebla2017probing,
	genway2014generalized,safavi2018verification,pedernales2015quantum,koch2023quantum,mei2022experimental,li2022observation}, superconducting qubits \cite{yoshihara2017superconducting,nataf2010no-go,zheng2023observation,zhao2022frustrated}, 
	quantum dots \cite{scheibner2007superradiance,yukalov2010dynamics,kozin2023quantum}, and 
	ultracold atoms \cite{hamner2014dicke,baumann2010dicke,baumann2011exploring,schmidt2014dynamical,landini2018formation,kroeze2018spinor,baden2014realization,hamner2014dicke,sauerwein2023engineering,zhang2021observation,hayashida2023perfect}, {\it etc. } 
	When $g_2 = 0$ and $g_1 =0$, it reduces to the Jaynes-Cummings (JC) model and anti-Jaynes-Cummings (aJC) model,
	respectively. Recent  investigations show that when $\omega/\Omega\rightarrow 0$, this model 
	exhibits a second-order phase transition from a normal phase to a superradiant phase \cite{hwang_quantum_2015, 
	puebla2017probing, liu_universal_2017, baumann2011exploring,zhuang2021universality,zheng2023observation}.
	In contrast to the models (A - C), Eq. \ref{eq-H} enables us to study the quantumness across the 
	critical point. Since the rotating and anti-rotating wave terms are ubiquitous, this model enables us to explore
	their individual role on quantumness.
	
	\begin{figure}
		\includegraphics[width=0.45\textwidth]{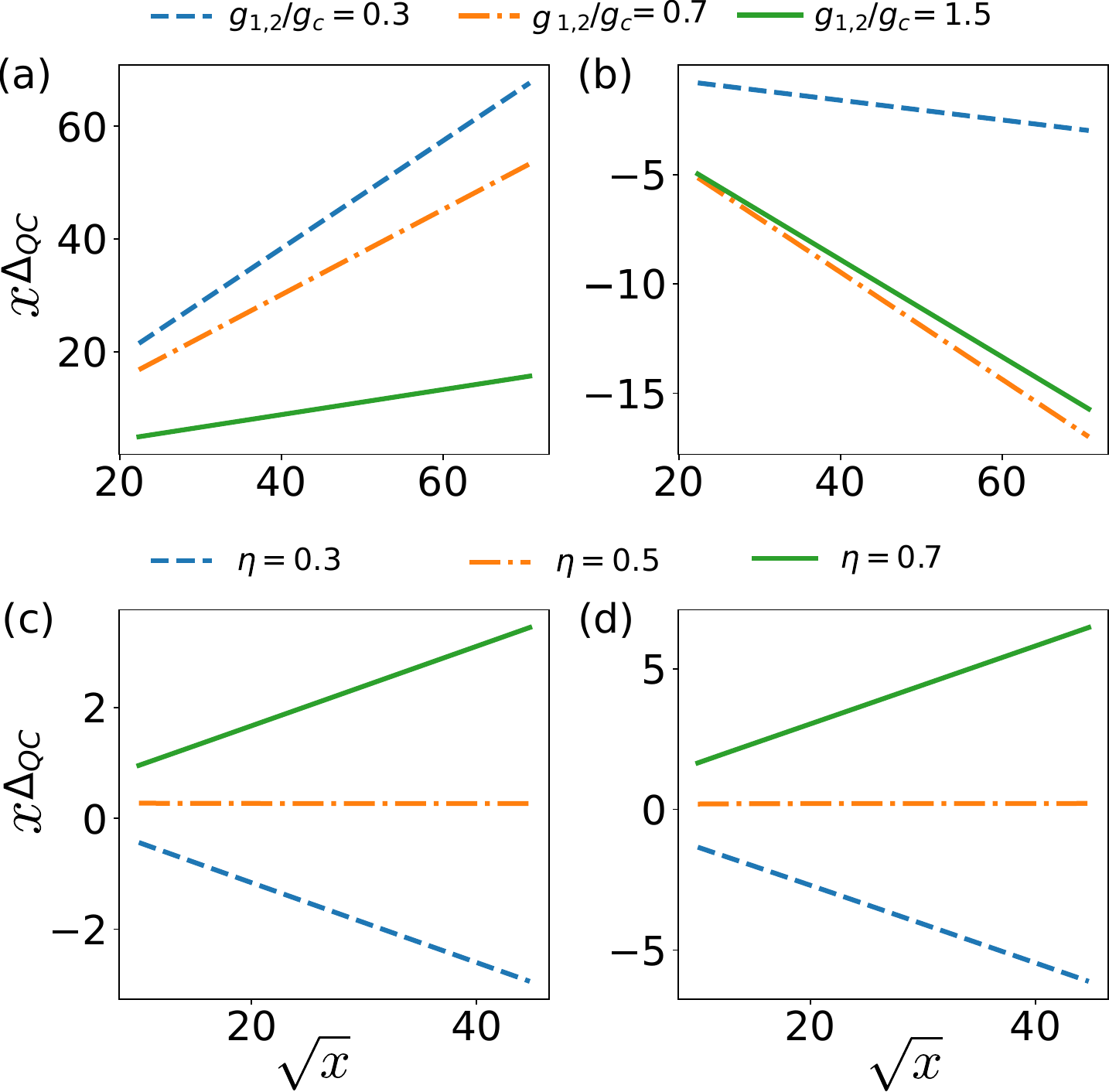}
		\caption{Verification of quantumness in Eq. \ref{eq-DeltaQC}. 
		(a) JC model with $g_2=0$, $g_1=0.3$, $0.7$, $1.5$, with $\beta=2$; 
		(b) aJC model with $g_1=0$, $g_2=0.3$, $0.7$, $1.5$, with $\beta=2$; 
		(c) Rabi model with $g=0.6$ (normal phase) and 
		(d) $g=1.2$ (superradiant phase), with $\beta=5$ and $\eta=g_1/g_2$. 
		In the Rabi model, $A=0$, hence $x \Delta_{\mathrm{QC}}$ is independent of $x$. 
		In all figures, we set $g_c =1$.}
		\label{fig-fig2}
	\end{figure}
	
	\begin{figure}
		\includegraphics[width=0.45\textwidth]{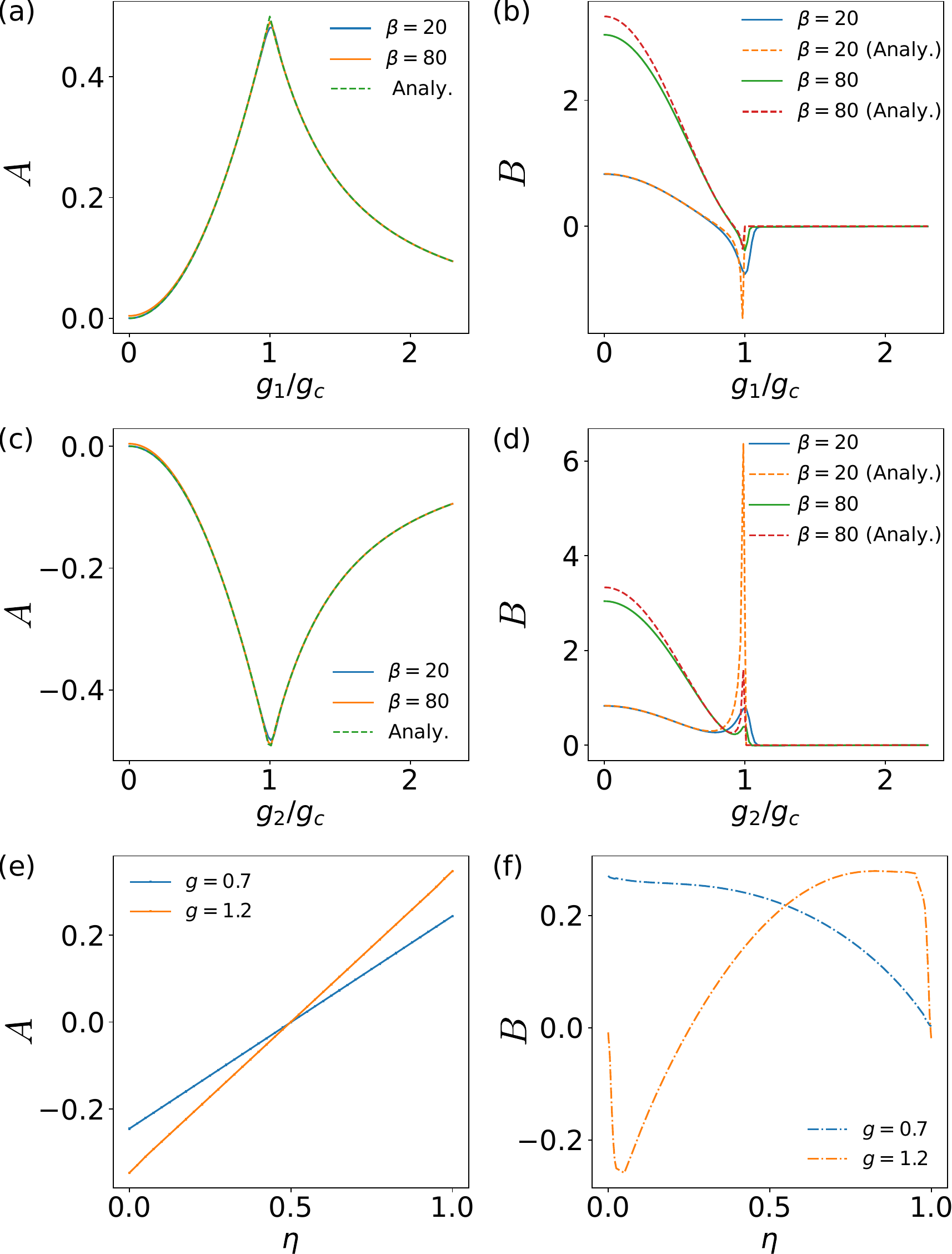}
		\caption{Coefficients of $A$ and $B$ in these three models. 
		The red lines in (a) - (d) are calcualted by numerial fitting of $\Delta_\text{QC}$ 
		at $\beta = 20$, $80$ for various $g_1$ and $g_2$, and the dashed lines are these 
		by the analytical expressions in Eq. \ref{eq-JC-AB} and Eq. \ref{eq-aJC-AB}. The 
		other parameters are: (a - b) $g_2 = 0$ for JC model; (c - d) $g_1 =0$ for the aJC 
		model; and (e -f) for the generalized Rabi model for $g < g_\text{c}$ and $g > g_\text{c}$.
		$\eta = 0$ and $\eta  = 1$ correspond to the analytic results of JC and aJC model.
		In (e) the two curves are identical to the empirical formula of Eq. \ref{eq-Aeta}.}
		\label{fig-fig3}
	\end{figure}
	
	\begin{table*}[]
		\centering
		\caption{ Summarized free energies in the normal phase (n) and superradiant (sr) phases. In the first column, JC, aJC refer to
			Jaynes-Cummings model and anti-Jaynes-Cummings model, respectively. The free energies are obtained from the partition 
			functions $Z_\text{C} = e^{-\beta F_\text{C}}$ and $Z_\text{Q} = e^{-\beta F_\text{Q}}$ using classical and quantum treatments, and in the classical model, JC and aJC have the same
			free energy, because in the classical level these models are the same (upon a transformation $a \rightarrow a^*$).}
		\begin{tabular}{|p{0.13\textwidth}|p{0.52\textwidth}|p{0.34\textwidth}|}
			\hline
			Model (n/sr) & Quantum free energy  $F_\text{Q}$ &      Classical free energy $F_\text{C}$             \\ \hline
			JC (n)	 &$-{\hbar g_\text{c} \sqrt{x} \over 2} - {1\over \beta} \ln ({\sqrt{x} \over \hbar g_\text{c} \beta (1-q^2)})
			+ {\hbar g_\text{c} q^2 \over 2\sqrt{x}} + {\hbar^2 \beta^2 g_\text{c}^2 (q^2-1)^3 + 24 q^2 \over 24\beta (q^2-1) x}$
			& \multirow{2}{*}{$-{\hbar g_\text{c} \sqrt{x} \over 2} - {1\over \beta} \ln ({\sqrt{x} \over \hbar g_\text{c}  \beta (1-q^2)})$}  \\  \cline{1-2}
			aJC (n)            & -$\frac{\hbar g_{c}\sqrt{x}}{2}-\frac{1}{\beta}\ln(\frac{\sqrt{x}}{\hbar \beta g_{c}(1-q^{2})})-\frac{\hbar g_{c}q^{2}}{2\sqrt{x}}+\frac{-\hbar ^2 \beta^2 g_{c}^{2}(q^{2}-1)^{3}+24q^{2}(2q^{2}+1)}{24x\beta(1-q^{2})}$    &                    \\ \hline
			JC (sr)		            & $-{\hbar g_\text{c} (x(q^4x + x-2) + 1) \over 4 q^2 x^{3/2}} - {1\over \beta} \ln (q \sqrt{{\pi \over \hbar \beta g_\text{c}}} x^{3/4})$	   & \multirow{2}{*}{$-{(1+q^4) \hbar g_\text{c} \over 4q^2} \sqrt{x} - {1\over \beta} \ln (q \sqrt{{\pi \over \beta \hbar g_\text{c}}} x^{3/4})$} \\ \cline{1-2}
			aJC (sr)            & $-{\hbar g_\text{c} 	(x(q^4x + x+2) + 1) \over 4 q^2 x^{3/2}} - {1\over \beta} \ln (q \sqrt{{\pi \over \hbar \beta g_\text{c}}} x^{3/4})$     &         \\ \hline
		\end{tabular}
		\label{tableI}  
	\end{table*}
	
	\begin{figure}
		\includegraphics[width=0.45\textwidth]{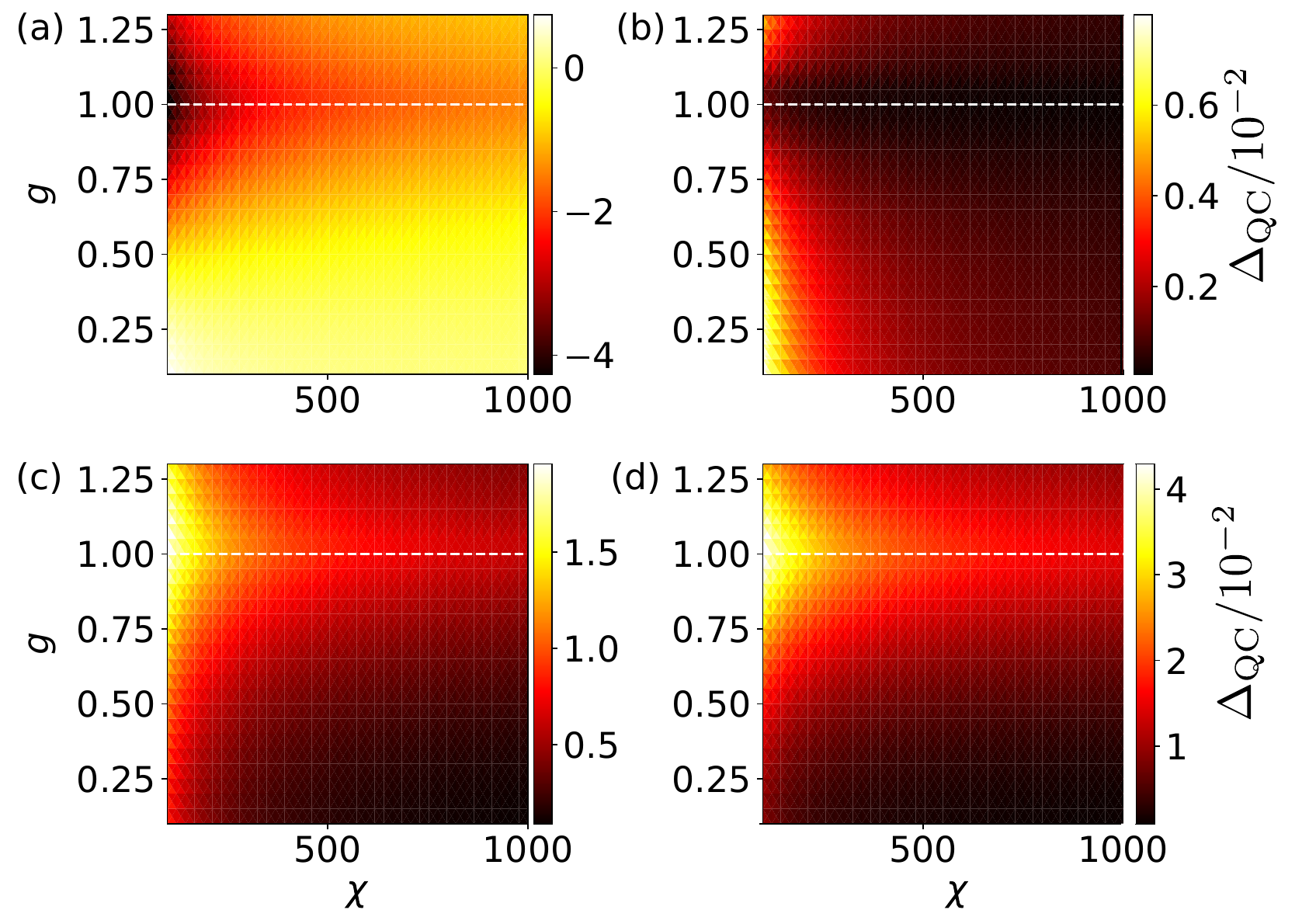}
		\caption{Quantumness $\Delta_\text{QC}$ near the critical point in the generalized Rabi model. 
			(a) $\eta = 0$ (for the JC model); 
			(b) $\eta = 0.5$ (for the ideal Rabi model); 
			(c) $\eta = 0.75$ (for the generalized Rabi model); 
			(d) $\eta = 1$ (for the aJC model).  In
			all figures, we have used $g_\text{c} = 1$ and $\beta = 5$.}
		\label{fig-fig4}
	\end{figure}
	
	(I). {\it JC model with $g_1 \ne 0$ and $g_2 = 0$}. Let $q = g_1/g_\text{c}$, $\omega = g_\text{c}/\sqrt{x}$, and $\Omega = g_\text{c} \sqrt{x}$, where $x = \Omega/\omega$, such that $\omega\Omega = g_c^2$ is fixed during the variation of the parameters. 
	This model can be solved analytically because the whole Hilbert space can be decoupled into small subspaces 
	(Fig.  \ref{fig-fig1} (a)), with eigenvalues $E_n^\pm = \hbar ( n + 1) \omega \pm  \hbar \sqrt{ g_1 ^2 (n + 1) + \frac{(\omega  - \Omega)^2}{4}}$ for $n \ge 0$. The configuration in Fig. \ref{fig-fig1} (a) is of particular 
	importance in the limit $\Omega \ll k_B T$, where only the lower branch $E_n^-$ is significant.
	The phase transition of this model can be manifested from the vanished energy gap, as shown in Fig. \ref{fig-fig1} (b); 
	and the corresponding spectra are shown in Fig. \ref{fig-fig1} (c). 
	In the  normal phase, we expect $E_n^- \propto n$; and in the superradiant phase, we expect $E_n^- \sim (n-n_c)^2$. In the
	phase boundary, we have $n_c =0$ and $E_n^- \propto n^2$. With this feature, the summation of $n$ can be generalized from 
	zero to infinity (see the filled regime in Fig.  \ref{fig-fig1} (c)). Using the Euler-Maclaurin formula, we can obtain the 
	free energy in these two phases.
	Meanwhile, the classical free energy is obtained by defining $a = \sqrt{\omega / 2\hbar }(\hat{x} + i\hat{p}/\omega )$ 
	and $a^\dagger = a^*$, which is commonly used  to
	analyze its phase transition in the limit $\omega/\Omega \rightarrow 0$ (see Eq. \ref{eq-harmonic})
	\cite{emary_quantum_2003, emary_chaos_2003,zhuang2021universality}. 
	Then the classical partition function is given by $Z_\text{C} = \sum_{s=\pm}\int dx dp/(2\pi \hbar) e^{-\beta E_{s}(a, a^*)}$, in which $E_{\pm} = \hbar \omega |a|^2 \pm \hbar \sqrt{(a^\ast g_1 + a g_2 )(a g_1 + a^\ast g_2  ) + \Omega^2}$. 
	We can solve this model using the saddle point method, which is almost exact in the low 
	temperature regime. The results of $F_\text{Q}$ and $F_\text{C}$
	for these phases are summarized in Table \ref{tableI}, yielding
\begin{equation}
		\Delta_\text{QC} = {A \over \sqrt{x}} + {B \over x} + \cdots, \quad x = {\Omega \over \omega}. 
		\label{eq-DeltaQC}
	\end{equation}
	If we write $\Delta_\text{QC}(\omega) = \Delta_\text{QC}'(0)\omega + \Delta_\text{QC}''(0)\omega^2/2$, then 
	$A \propto \Delta_\text{QC}'(0)$ and $B \propto \Delta_\text{QC}''(0)$. Thus QCT happens when 
	$x\rightarrow \infty$. We find analytically that
	\begin{align}
		\begin{split}
			A=\begin{cases}
				\frac{\hbar g_{c}q^{2}}{2} \\
				\frac{\hbar g_{c}}{2q^{2}}
			\end{cases},
			B = \begin{cases}
				\frac{\hbar^2 \beta^2 g_\text{c}^2 (q^2 -1 )^3  + 24 q^2 }{24 \beta (q^2 -1 )} & q < 1\\
				0  & q > 1
			\end{cases}. {\tiny }
		\end{split}
		\label{eq-JC-AB}
	\end{align}
	
	(II). {\it aJC model with $g_2 \ne 0$ and $g_1 = 0$}. We define
	$q = g_2/g_\text{c}$, while the other parameters are the same as (I). This
	model can also be calculated exactly using $	E_{n}^{\pm} =
	(n+\frac{1}{2})\hbar \omega \pm \frac{1}{2}( \hbar \omega-\sqrt{4 \hbar^2 g^{2}(n+1)+\hbar^2 (\omega+ \Omega)^{2}})$
	for $n \ge -1$, which is slightly different from that in the JC model. The results are summarized in Table 
	\ref{tableI}, with quantumness the same as Eq. \ref{eq-DeltaQC}. We have 
	\begin{align}
		\begin{split}
			A = \begin{cases}
				-\frac{\hbar g_{\text{c}}q^{2}}{2} \\
				-\frac{\hbar g_{\text{c}}}{2 q^{2}}
			\end{cases}, 
			B = \begin{cases}
				\frac{\hbar^2 \beta^2 g_\text{c}^2 (q^2 -1 )^3  - 24 (2q^2 + q^4) }{24 \beta (q^2 -1 )} & q < 1\\
				0  & q > 1
			\end{cases}.
		\end{split}
		\label{eq-aJC-AB}
	\end{align}
	The classical energies are the same (see Table \ref{tableI}).
	Physically, quantumness is fully determined the discrete nature of quantum mechanics.
	For example, when $\omega \ll \Omega$, we have $E_n^-(\omega) = E_n^-(0) + \kappa \omega$, 
	where $\kappa = (dE_n^-(\omega)/d\omega)$. When $\kappa \ne 0$, we naturally have the $A$ 
	term. Therefore, when all these terms disappear, quantumness  naturally is  vanishes, indicating QCT.

	The results in Eqs. \ref{eq-JC-AB} - \ref{eq-aJC-AB} already reveal some
	important and sutble difference between the JC and aJC
	models. The scaling law of Eq. \ref{eq-DeltaQC} for these two cases
	are shown in Fig. \ref{fig-fig2} (a) and (b); and the corresponding
	coefficients of $A$ and $B$ are shown in Fig. \ref{fig-fig3}  (a) to (d).
	In both models, the term $A$ is always important in the
	whole phase regime. Especially, $A$ will takes on a maximal value at the phase
	boundary. We find that $A$ in the JC and aJC models take opposite values, 
	showing that the signs of the quantumness in these two models are different. The negative 
	quantumness is in stark contrast to that in the harmonic oscillator (Eq. \ref{eq-harmonic}).
	Besides, we find that the coefficient $B$ is only important in the normal phase, while in the 
	supperradiant phase, it becomes vanished. 
	
	(III). {\it Generalized Rabi model with $g_1 g_2 \neq 0$}. Let $\eta= g_1 /(g_1+g_2)$, the 
	phase transition happens at $(g_1+g_2)^2=\omega \Omega$. We only need to consider the regime 
	with $g_i\ge 0$ \cite{g1g2sign}. When $\eta=1 / 2$, it yields the Rabi model. This model can not be solved 
	analytically, and we solve this model using numerical method. We find that the quantumness can still be written 
	in the general form as Eq. \ref{eq-DeltaQC}. Our numerical results of quantumness are presented 
	in Fig.  \ref{fig-fig2} (c) - (d) and the extracted coefficients $A$ and $B$ 
	are shown in Fig. \ref{fig-fig3} (e) and (f). Strikingly, we find that when $\eta=1/2$, 
	the quantumness is only given by the $B$ term, with $A=0$. Thus due to the counteract effect of the JC and aJC 
	interactions, the Rabi model has the smallest quantumness. From Fig. \ref{fig-fig3} (e)  we have 
	\begin{equation}
		A(\eta) = A(0)(1-\eta) + A(1) \eta, 
		\label{eq-Aeta}
	\end{equation}
	where $A(0) = - A(1)$ are the coefficients in the aJC and JC models, respectively. This relation 
	naturally explains the vanished $A$ in the Rabi model.

	\begin{figure}
		\includegraphics[width=0.45\textwidth]{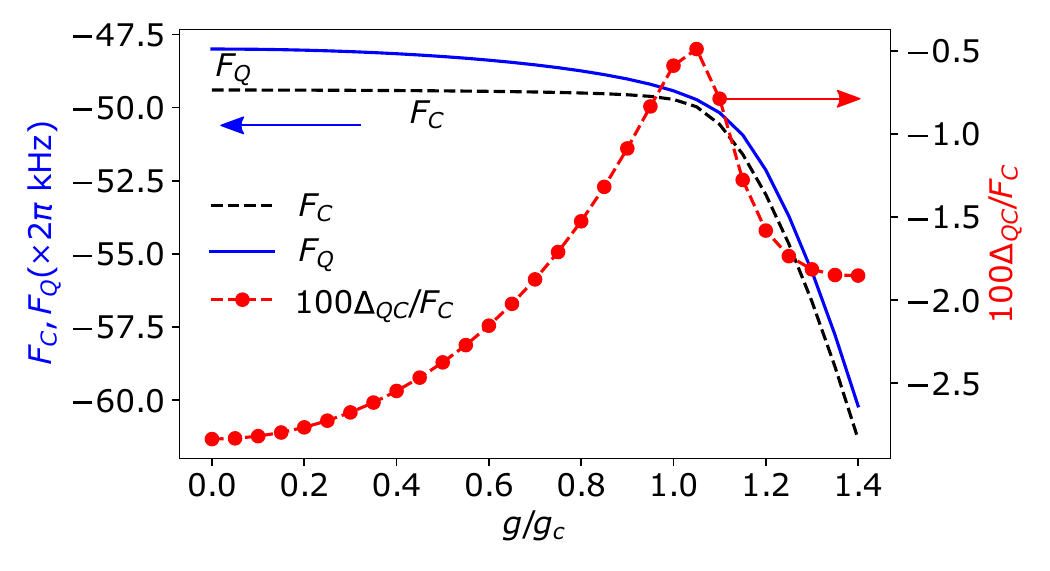}
		\caption{QCT in experimental Rabi model with trapped ions \cite{cai2021observation}. Here we have used 
		$\omega=2 \pi \times 4.0$ kHz, $ \Omega=2 \pi \times 100.0 $ kHz, which corresponds to a ratio 
		$x=25$, and the critical point is at $g_\text{c}=2 \pi \times 20.0 $ kHz, showing that 
		$\Delta_\text{QC}/F_\text{C} \sim 1\%$. 
		}
		\label{fig-fig5}
	\end{figure}
	
	(IV). {\it Quantumness and phase transition}: 
	It is natural to explore the quantumness across the critical point, which has 
	the strongest fluctuations in both classical and quantum mechanics. However, their difference 
	is not necessary to be increased. From the sign of $A$, we see that the contribution of JC 
	and aJC are opposite, which can yield totally different behaviors. The numerical 
	results for these models are presented in Fig. \ref{fig-fig4}. We find that with the 
	increase of $g$, the 
	quantumness is enhanced in the JC and aJC models; however, in the Rabi model, it is suppressed 
	near the critical point. These results 
	suggest that the quantumness is an intrinsic  property of the Hamiltonian, which can be increased 
	or decreased by the critical fluctuations. As a result, it is only in this limit $\Delta_\text{QC} =0$ that 
	the classical phase transition can happen in the general Rabi model \cite{zhuang2021universality}.

	{\it Quantumness in experiment with trapped ions}: Finally, we discuss the  relevance of our results in the current experiments with trapped ions implemented in the experiments in Duan group \cite{cai2021observation}. In this experiment with Rabi model, $\omega=$ $2 \pi \times 4.0 $ kHz, $\Omega=2 \pi \times 100.0  $ kHz, $x=25$, and the critical point is at $g_\text{c}=2 \pi \times 20.0 $ kHz (at $T=0$). By varying of the coupling $g$ from 0 to $14 $ kHz $(1.4g_\text{c})$, we estimate $F_\text{Q}, F_\text{C}$ and $\Delta_\text{QC}$ in these experiments in Fig. \ref{fig-fig5} , showing that in the current setup, the quantumness is about $1 \%$ of the total classical energy. This value is small, yet nonzero, 
thus in experiment only a smooth crossover between the two phases can be demonstrated. We find that to observe a 
relative sharp transition, $x$ must be increased to $100$, yielding $\Delta_{\mathrm{QC}} / F_\text{C}$ to be further 
reduced by one order of magnitude since $\Delta_{\mathrm{QC}} / F_\text{C} \sim x^{-3 / 2}$, for $A=0$ 
and $F_\text{C} \propto \sqrt{x}$; see Table \ref{tableI}.

	To conclude, we define the quantumness $\Delta_\text{QC}$ to quantitatively characterize the QCT, and examine this 
	quality in the generalized Rabi model. This interpretation does not 
	rely on the state collapse, thus is not a feature of some particular states 
	as discussed in quantum information science \cite{Modi2012, zurek2003decoherence, schlosshauer2014quantum}, instead, 
	it is a feature of the Hamiltonian in thermal equilibrium. We show explicitly
	that not only the temperature and Planck constant, but also the system parameters are important for the QCT. This 
	quantity can be applied to study the QCT in a broad range of many-body physical models, and clarify the roles of quantumness 
	on observations and phase transitions, which could help to resolve the long-sought mysterious boundary between quantum and 
	classical worlds. From this perspective, we expect this quantity to be important for these models which can not 
	be described approximately by classical mechanics. It also has potential to be unified with the 
	decoherence theory for QCT 
	\cite{zurek1991decoherence, Haroche_1998, Zeh1970, paz2002environment,schlosshauer2014quantum,
    zeh1970interpretation,zurek1981pointer,schlosshauer2005decoherence,zurek1982environment,zurek2003decoherence}.

\textit{Acknowledgements}: This work is supported by the National Natural Science Foundation of China (NSFC) with No. 11774328, 
No. 11834005, No. 12247101, the Innovation Program for Quantum Science and Technology (No. 2021ZD0301200 and No. 2021ZD0301500),
and National Key Research and Development Program of China (Grant No. 2022YFA1402704).

\bibliography{ref}

\begin{thebibliography}{60}
\bibitem [{g1g()}]{g1g2sign}%
  \BibitemOpen
  \href@noop {} {}\bibinfo {note} {We focus on $g_1 > 0$ and $g_2 > 0$. When
  these two coefficients have opposite sign, we can make a transformation $a
  \rightarrow i a$, and $\sigma^- \rightarrow i \sigma^-$ to change them to
  have the same sign. When $g_1 < 0$ and $g_2 < 0$, we can change $a
  \rightarrow - a$ to change them to be positive valued.}\BibitemShut {Stop}%
\end{thebibliography}
	
\end{document}